%%%%%%%%%%%%%%%%%%%%%%%%%%%%%%%%%%%%%%%%%%%%%%%%%%%%%%%%%%%%%%%%%%%%%%%%%%
%                                                                        %
%                 This is typed in Plain TeX,no figures                  %
%                                                                        %
%%%%%%%%%%%%%%%%%%%%%%%%%%%%%%%%%%%%%%%%%%%%%%%%%%%%%%%%%%%%%%%%%%%%%%%%%%

\font\twelvebf=cmbx12
\font\ninerm=cmr9
\nopagenumbers
\magnification =\magstep 1
\overfullrule=0pt
\baselineskip=18pt
\line{\hfil CCNY-HEP 96/10}
\line{\hfil hep-th 96mmnnn}
\line{\hfil August 1996}
\vskip .8in
\centerline{\twelvebf On Non-Abelian Thomas-Fermi Screening} 
\vskip .5in
\centerline{\ninerm G. ALEXANIAN and V.P. NAIR \footnote {*}{E-mail:
garnik@scisun.sci.ccny.cuny.edu, vpn@ajanta.sci.ccny.cuny.edu}}
\vskip .1in
\centerline{ Physics Department}
\centerline{City College of the City University of New York}
\centerline{New York, New York 10031}
\vskip 1in
\baselineskip=16pt
\centerline{\bf Abstract}
\vskip .1in
The Thomas-Fermi screening of non-Abelian gauge fields by fermions
or screening of gluon fields in quark matter is discussed. It is described
by an effective mass term which is, as with hard thermal loops, related to
the eikonal for a Chern-Simons theory and the Wess-Zumino-Witten action.
\vfill\eject
\footline={\hss\tenrm\folio\hss}

\def \12 {{\textstyle {1\over 2}}}

	The screening of electromagnetic interactions in a plasma
of charged particles, the so-called Debye screening, is quite well-known. 
An analogous  effect, the Thomas-Fermi screening, occurs in
the case of a degenerate charged fermion system such as the
electron gas in metal [1]. Both effects can be easily understood, in
a field-theoretic language, by calculating the one loop photon
propagator in which the charged particle  propagators are at finite
temperature and density. (The two effects are in fact quite similar;
for the degenerate fermion gas, the excitations generated by the
propagating photon are particle-hole pairs which behave like plasma
background). The non-Abelian analogues of these screening effects are
of considerable interest especially in view of the possibility of
obtaining hot and dense quark matter systems in heavy ion collisions [2].
A non-Abelian Thomas-Fermi effect can also be of interest in
calculations of the equation of state for quark matter inside neutron
stars. Non-Abelian Debye screening and related effects have been
intensively investigated over the last few years [3-8]. It is well-known
that a proper calculation of this effect involves summing up all the
hard thermal loop Feynman diagrams [3]. This results in a gauge-invariant,
nonlocal,effective mass term for the gauge bosons (or gluons in a
chromodynamic context). This mass term has many nice properties being
closely related to Chern-Simons and Wess-Zumino-Witten (WZW) theories [5,6].

In this letter we analyze the non-Abelian Thomas-Fermi screening
for degenerate quark matter of finite baryon number. Since the quark
contribution to the two-point function for gluons is similar to that
for photons, there should be Thomas-Fermi screening for quark matter.
For reasons of non-Abelian gauge invariance, as with Debye screening,
there will be higher point contributions, the whole series again
summing up to an effective mass term. This term will have the
same structure as the effective action for hard thermal loops; the
numerical value of the screening mass, however, will be determined
by the chemical potential rather than temperature.

Consideration of the one loop two-point function shows
that the screening mass $\sim g\mu$ where $g$ is the coupling
constant and $\mu$ is the chemical potential. It is then clear
that a higher loop diagram in which such a term is inserted  can give
contributions of the same order for the integration range of loop
momenta ${\buildrel < \over \sim}g\mu$. This is exactly as in the hard thermal 
loop case. One must therefore sum up diagrams with loop momenta
$\sim \mu$ and external momenta ${\buildrel < \over \sim} g\mu$. This effective
action must then be used for a self-consistent evaluation of
the screeening mass. We obtain this effective action in what
follows.

	Let us start by considering the two-point function for
gluons. The conserved baryon charge is given in terms of the
quark field $q(x)$ by $\int q^\dagger q $. With a chemical potential
term $\mu\int q^\dagger q$ added to the action, the quark propagator
is given by
$$
\eqalign{
S(x,y)=\langle T q(x)\bar q(y)\rangle=
\int {d^3p\over {(2\pi)}^3}{1\over 2p^0}\bigl[
\{&\theta(x^0-y^0)\alpha_p\gamma\cdot p e^{-ip(x-y)}+
\bar\beta_p\gamma\cdot p'e^{ip'(x-y)}\} +\cr
-&\theta(y^0-x^0)\{\beta_p\gamma\cdot p e^{-ip(x-y)}+\bar\alpha_p
\gamma\cdot p' e^{ip'(x-y)}\}\bigr]
}
\eqno(1)
$$
where $p^0=|\vec p|, ~p=(p^0,\vec p),~ p'=(p^0,-\vec p)$ and $\theta(x)$ is
the step function. Also
$$
\alpha_p=1-n_p, ~~~~~~~~~~\beta_p=n_p
\eqno(2)
$$
The distribution functions $n_p,{\bar n}_p$ corresponding to quarks
and antiquarks respectively are given by
$$ n_p={1\over e^{(p^0-\mu) / T}+1},~~~~~~~~ \bar n_p={1\over e^{(p^0+\mu)
 / T} +1}
\eqno(3)
$$
The one-loop quark graphs are given by the effective action
$$
\Gamma=-i{\rm Tr} ~\log (1+S\gamma\cdot A)
\eqno(4)
$$
$A_\mu=-it^a g A^a_\mu$ is the gluon vector potential, $t^a$
are hermitian matrices corresponding to the generators of the
Lie algebra in the quark representation.
In the above expression for $\Gamma$ a functional trace is implied
as well as the trace over the spin and color labels.
The two-gluon term in
$\Gamma$ is given by
$$
\Gamma^{(2)}={i\over 2}\int d^4x~d^4y ~{\rm Tr}\bigl[\gamma\cdot A(x)
S(x,y) \gamma\cdot A(y) S(y,x)\bigr]
\eqno(5)
$$
Using equation (1) and carrying out the time-integrations we get
$$
\eqalign{
\Gamma^{(2)}=-{1\over 2}\int d\mu(k) {d^3q\over{(2\pi)}^3}{1\over 2p^0}
{1\over 2q^0}\bigl[&T(p,q)\bigl({\alpha_p\beta_q\over p^0-q^0-k^0-i\epsilon}
-{\alpha_q\beta_p\over p^0-q^0-k^0+i\epsilon}\bigr)+\cr
&T(p,q')\bigl({\alpha_p\bar\alpha_q\over p^0+q^0-k^0-i\epsilon}-
{\beta_p\bar\beta_p\over p^0+q^0-k^0+i\epsilon}\bigr)+\cr
&T(p',q)\bigl({\bar\alpha_p\alpha_q\over p^0+q^0+k^0-i\epsilon}-
{\bar\beta_p\beta_q\over p^0+q^0+k^0+i\epsilon}\bigr)+\cr
&T(p',q')\bigl({\bar\alpha_p\bar\beta_q\over p^0-q^0+k^0-i\epsilon}
-{\bar\beta_p\bar\alpha_p\over p^0-q^0+k^0+i\epsilon}\bigr)
\bigr]
}
\eqno(6)
$$
where
$$
\eqalign{
A_\mu (x)&=\int{d^4k\over{(2\pi)}^4} e^{ikx}A_\mu(k)\cr
d\mu(k)&={(2\pi)}^4\delta^{(4)}(k+k'){d^4k\over{(2\pi)}^4}
{d^4k'\over{(2\pi)}^4}\cr
T(p,q)&=Tr\bigl[\gamma\cdot A(k)~\gamma\cdot p~\gamma\cdot A(k')~\gamma\cdot 
q\bigr]
}
\eqno(7)
$$
and $\vec p=\vec q+\vec k$ in equation(6).

	The $i\epsilon's$ can be taken to go to zero at this stage.
They were introduced for convergence of time-integrations and
contribute to the imaginary part. Here we are interested in
screening effects which are described by the real part of the
two-point function. (Also for many physical situations, the relevant
imaginary part is that of the retarded function which is not directly given
by the above time ordered function [6].) Further we are interested in a degenerate
gas of quarks; it is therefore appropriate to consider $T\ll\mu$.
As $T/\mu \rightarrow 0$, the antiquark occupation numbers 
$\bar n_p\rightarrow 0$
(for positive $\mu$). Equation (6) then simplifies as
$$
\eqalign{
\Gamma^{(2)}=-{1\over 2}\int d\mu(k) {d^3q\over{(2\pi)}^3}{1\over 2p^0}
{1\over 2q^0}\bigl[&T(p,q)
{(n_q-n_p)\over p^0-q^0-k^0}+
T(p,q'){n_p\over p^0+q^0-k^0}\cr-
&T(p',q){n_q\over p^0+q^0+k^0}
\bigr]
}
\eqno(8)
$$
As explained in  the introduction, the relevant kinematic
regime is $|\vec p|, |\vec q| \gg |\vec k|$, so that $p^0-q^0-k^0\approx -k\cdot Q$,
$p^0+q^0\pm k^0\approx 2q^0, Q=(1,{\vec q / q^0})$. Further,
$$
\eqalign{
&T(p,q)\approx 8{q^0}^2 tr(A_1\cdot Q A_2\cdot Q)\cr
&T(p',q)\approx T(p,q')\approx 4{q^0}^2
tr(A_1\cdot Q' A_2\cdot Q + A_1\cdot Q A_2\cdot Q'
-2A_1\cdot A_2)
}
\eqno(9)
$$
where $ A_1=A(k), ~A_2=A(k'),~Q'=(1,-{\vec q / q^0})$.
Expression (8) now simplifies as
$$
\eqalign{
\Gamma^{(2)}=-{1\over 2}\int d\mu(k) {d^3q\over{(2\pi)}^3}
{\rm tr}\bigl[
{dn\over dq^0} &{A_1\cdot Q A_2\cdot Q\over k\cdot Q}2\vec Q\cdot \vec k
-{n\over q^0}(A_1\cdot Q' A_2\cdot Q +\cr
&A_1\cdot Q A_2\cdot Q'-2A_1\cdot A_2)
\bigr]
}
\eqno(10)
$$
Using $\int d^3q~ {dn\over dq^0}f(Q)=-\int d^3q~{2n\over q^0}f(Q)$,
and some properties of $\vec Q$-integration, $\Gamma^{(2)}$
simplifies to
$$
\eqalign{
\Gamma^{(2)}=&-{1\over 2}\int d\mu(k)\int {d^3q\over{(2\pi)^3}}
~{n\over 2q^0}~ 8~{\rm tr}\bigl[2A_{1+}A_{2-}-{k\cdot Q'\over k\cdot 
Q}A_{1+}A_{2+} -{k\cdot Q\over k\cdot Q'}A_{1-}A_{2-}\bigr]\cr
=&-{1\over 2}\int d\mu(k)\Bigl({\mu^2\over 4\pi^3}\Bigr)\int d\Omega~
{\rm tr}\bigl[2A_{1+}A_{2-}-{k\cdot Q'\over k\cdot Q}A_{1+}A_{2+}
-{k\cdot Q\over k\cdot Q'}A_{1-}A_{2-}\bigr]
}
\eqno(11)
$$
where $A_+={A\cdot Q\over 2}, A_-={A\cdot Q'\over 2}$
and we have taken the limit of $T\rightarrow 0$ (small compared to $\mu$).
As with hard thermal loops, we can transform this back to coordinate
space and write
$$
\Gamma^{(2)}=-{\mu^2\over 4\pi^3}\int d\Omega ~K^{(2)}(A_+,A_-)
\eqno(12)
$$
where
$$
\eqalign{
&K(A_+,A_-)=\bigl[\int d^4x~{\rm tr}(A_+A_-)+i\pi I(A_+)+i\pi \tilde I(A_-)
\bigr]\cr
&I(A_+)=i\sum_2^\infty{(-1)^n\over n}\int d^2x^T
{d^2z_1\over\pi}\cdots{d^2z_n\over\pi}
{tr[A_+(x_1)\cdots A_+(x_n)]\over
(\bar z_1-\bar z_2)\cdots(\bar z_n-\bar z_1)}
}
\eqno(13)
$$
$\int d\Omega$ defines integration over the orientations of $\vec Q$.
$K^{(2)}$ in equation (12) denotes the terms
in $K$ which are quadratic in $A's$; $z$ and $\bar z$ are the Wick-rotated
versions of $x\cdot Q'$ and $x\cdot Q$ respectively and $x^T$ is
transverse to $\vec Q$, i.e $\vec Q\cdot{\vec x}^T=0$. ${\tilde I}(A_{-})$ is
obtained from $I(A_+)$ by $z \leftrightarrow {\bar z}$.
 $I(A_+)$, as
explained elsewhere, is essentially the eikonal function for Chern-Simons
theory. $K(A_+,A_-)$ can be related to the
WZW-action for a hermitian matrix $M^\dagger M$ defined in terms of $A_\pm$.

Consider now the three-point function.
One has twenty-four terms
with denominators involving different combinations of loop
momenta and external momenta.
The simplification of this expression proceeds in much the same
way as for $\Gamma^{(2)}$ - one can 
neglect terms with
denominators that are of the order of $\mu$ and leave
the terms that involve differences between loop momenta, which
are of the order of $k\ll\mu$.
After that, the relevant contributions will be
$$
\eqalign{
\Gamma^{(3)}=\int {d^3q\over (2\pi)^3}\,2\,
\Bigl\{ &{n_p\over k_2^0+p^0-q^0}{1\over r^0-q^0-k_3^0}+
{n_r\over q^0-r^0+k_3^0}{1\over p^0-r^0-k_1^0}+\cr
&{n_p\over p^0-r^0-k_1^0}{1\over p^0-q^0+k_2^0}
\Bigr\}~{\rm Tr}(A_1\cdot QA_2\cdot QA_3\cdot Q)
}
\eqno(14)
$$
where $p,q,r$ are the loop momenta, $k_1,k_2$ and $k_3$ are the
momenta of external gluons and $\vec p =\vec q - {\vec k}_2,
\vec r=\vec q +{\vec k}_3$. 
After performing the $dq^0$-integral the final expression is
$$
\Gamma^{(3)}={\mu^2\over (2\pi)^3} \int d\Omega \Bigl( {k_2\cdot Q'-k_2\cdot Q
\over k_1\cdot Q k_2\cdot Q} + {k_3\cdot Q - k_3\cdot Q'\over
k_3\cdot Q k_1\cdot Q}\Bigr)~{\rm Tr}(A_1\cdot QA_2\cdot
QA_3\cdot Q)
\eqno(15)
$$
Using definition (13) one can show that
$$
\Gamma^{(3)}=-{\mu^2\over 4\pi^3} \int d\Omega ~K^{(3)}(A_+,A_-)
\eqno(16)
$$
The fact that the same coefficient appears in both (12) and (14)
is crucial; this guarantees gauge invariance of the full 
effective action $\Gamma$.
The non-Abelian gauge-invariant completion of $K^{(2)}+K^{(3)}$ is given by
the full $K$ of equation (13). The final answer is thus
$$
\Gamma =-{\mu^2\over 4\pi^3}\int d\Omega ~ K(A_+,A_-)
\eqno(17)
$$

We now turn to the question of the coefficient of $K$ in equation (17).
In calculating the screening effects using $\Gamma$, we should not change the
Lagrangian for the theory, which may be, say, chromodynamics; we should only
rearrange terms in the perturbative expansion and sum certain classes
of diagrams. There are then two different but closely related ways of 
proceeding. We write the action as
$$
\eqalign{
S&=S_0+\Delta\int d\Omega ~K\cr
S_0&=S_{QCD}-m^2\int d\Omega ~K
}
\eqno(18)
$$
$S_0$ is used to define propagators and vertices and $\Delta$ is treated as
being nominally one loop order higher than $S_0$. In the first case, we
use the value as calculated above ( or the
analogous value for hard thermal loops) for $m^2$. After calculating
higher order corrections,
$\Delta$ is set to $m^2$ ( so that $S\rightarrow  S_{QCD}$ ). The corrections
are in general non-vanishing and this is useful
if the corrections are small compared to the lowest order value.
The alternative is to keep
$m^2$ as an arbitrary parameter and choose $\Delta$ ( as a function of $m$, say
$\Delta(m)$ ) so as to cancel
out all the corrections. Upon setting $\Delta$ to $m^2$, we get a gap
equation, $\Delta(m) = m^2$, which can be solved for $m$.
In the case of the hard thermal loops, the first approach is satisfactory.
The relevant distribution for the momentum-integration is
$$
f(q)dq={q^2dq\over e^{q/T} + 1}
\eqno(19)
$$
The lowest order calculation only evaluates the contribution from
the region of $q {\buildrel > \over \sim} T $.
The probability contained in the region
$q \geq {1\over 2}T$ for the distribution 
is approximately 0.99.
We can therefore expect that the neglect of the low $q$-regime is
not very significant for the numerical value of the screening mass ( 
within a calculational scheme with resummations
as explained above ). For the case of Thomas-Fermi screening,
the relevant distribution is
$$
f(q)dq\approx\theta(\mu-q)q^2dq
\eqno(20)
$$
The probability contained in the region $q\geq{1\over 2}\mu$
is now 0.875. We thus expect that the
lowest order value of the coefficient of
$-\int d\Omega ~K$ in equation (15), viz., $\mu^2/4\pi^3$, is somewhat less
accurate than the analogous quantity for hard thermal loops.
In this case a self-consistent calculational scheme for including
higher order effects might be more appropriate.

After this work was completed, we became aware of a paper by 
Cristina Manuel
where non-Abelian Thomas-Fermi screening is discussed in a kinetic
theory framework and results similar to ours are obtained [9].
Our approach, based on evaluation of Feynman diagrams, is
complementary to this work. We thank Cristina Manuel for bringing
this work to our attention.

This work was supported in part by the National Science Foundation
grant number PHY-9322591 and PSC-CUNY Research Award 667447. VPN also
thanks Professor N.Khuri for hospitality at Rockefeller University
where part of this work was done.
\vskip .1in
\noindent{\bf References}
\item{1.} See, for example, A.L.Fetter and J.D.Walecka, "Quantum theory
of many particle systems", McGraw-Hill, 1971.
\item{2.} See, for example, C.P.Singh, {\it Phys.Rep.} 
{\bf 236} (1993) 147. 
\item{3.} R.Pisarski, {\it Physica A} {\bf 158} (1989)
246; {\it Phys.Rev.Lett.} {\bf 63} (1989) 1129;
E.Braaten and R.Pisarski, {\it Phys.Rev.} {\bf D42} (1990)
2156; {\it Nucl.Phys.} {\bf B337} (1990) 569; {\it
ibid.} {\bf B339} (1990) 310; {\it Phys.Rev.} {\bf
D45} (1992) 1827.
\item{4.} J.Frenkel and J.C.Taylor, {\it Nucl.Phys.}
{\bf B334} (1990) 199; J.C.Taylor and S.M.H.Wong, {\it
Nucl.Phys.} {\bf B346} (1990) 115.
\item{5.} R.Efraty and V.P.Nair, {\it Phys.Rev.Lett.}
{\bf 68} (1992) 2891; {\it Phys.Rev.} {\bf D47} (1993)
5601.
\item{6.}  R.Jackiw and V.P.Nair, {\it Phys.Rev.} 
{\bf D48} (1993) 4991.
\item{7.} J.P.Blaizot and E.Iancu, {\it Phys.Rev.Lett.}
 {\bf 70} (1993) 3376;
{\it Nucl.Phys.} {\bf B 417} (1994) 608.
\item{8.} R.Jackiw, Q.Liu
and C.Lucchesi, {\it Phys.Rev.} {\bf D49} (1994) 6787;
P.F.Kelly {\it et al}, {\it Phys. Rev. Lett.}
{\bf 72} (1994) 3461; {\it Phys.Rev.} 
{\bf D50} (1994) 4209; V.P.Nair, {\it Phys.Rev.} 
{\bf D48} (1993) 3432;
{\it ibid.} {\bf D50} (1994) 4201; A.K.Rebhan, {\it Phys.Rev.}
{\bf D} (1993) 3967, {\it Nucl.Phys.} {\bf B 430} (1994) 319.
\item{9.} C. Manuel, {\it Phys.Rev.} {\bf D53} (1996) 5866.

\end